\documentclass[twoside,fleqn]{ActaStyle}
\usepackage{times}

\usepackage{lscape, amsbsy}
\usepackage{graphicx,epsfig,psfrag}
\usepackage[tbtags]{amsmath}

\newcommand{\dd}{\mathrm{d}}

\newcommand{\ba}{\begin{eqnarray}}
\newcommand{\ea}{\end{eqnarray}}
\newcommand{\bea}{\begin{array}{l}}
\newcommand{\eea}{\end{array}}
\newcommand{\ds}{\displaystyle}
\newcommand{\ldl}{\Lambda \partial_{\Lambda}}
\newcommand{\ie}{{\it i.e.}\ }
\newcommand{\eg}{{\it e.g.}\ }

\newcommand{\aka}{{\it a.k.a.}\ }
\newcommand{\ug}{\! = \!  }
\newcommand{\tr}{\mathrm{tr}}
\newcommand{\Lam}{\Lambda}
\newcommand{\demu}{\partial_\mu}
\newcommand{\denu}{\partial_\nu}
\newcommand{\C}{{\cal C}}
\newcommand{\A}{{\cal A}}
\newcommand{\hS}{{\hat S}}
\newcommand{\sh}{{\hat S}}
\def\eq#1{eq.~(\ref{#1})}
\def\sqr#1#2{{\vcenter{\vbox{\hrule height .#2pt
        \hbox{\vrule width .#2pt height#1pt \kern#1pt
              \vrule width.#2pt}
          \hrule height .#2pt}}}}
\def\Box{\sqr{6}{6}}
\def\hepth#1{{\tt hep-th/#1}}

\def\authorheading{S. Arnone et al.}

\def\shorttitle{Gauge invariant calculations}

\begin{document}

\pagerange{1}{8}

\title{MANIFESTLY GAUGE INVARIANT COMPUTATIONS}
\author{
Stefano Arnone\email{S.Arnone@soton.ac.uk}, 
Antonio Gatti\email{A.Gatti@soton.ac.uk}, 
Tim R. Morris\email{T.R.Morris@soton.ac.uk}
}
{
Department of Physics and Astronomy, University of Southampton\\ 
Highfield, Southampton SO17 1BJ, United Kingdom.
}

\day{May 14, 2002}

\abstract{Using a gauge invariant exact renormalization group, we show
 how to compute the effective action, and extract the physics, whilst
 manifestly preserving gauge invariance at each and every step. As an
 example we give an elegant computation of the one-loop $SU(N)$
 Yang-Mills beta function, for the first time at finite $N$ without any 
gauge fixing
 or ghosts. It is  also completely independent of the details put in by
 hand, \eg the choice of covariantisation and the cutoff profile, and,
 therefore, guides us to a procedure for streamlined  calculations.}
\section{Introduction}
\label{sec:intr} \setcounter{section}{1}\setcounter{equation}{0}
The continuum formulation of the Wilsonian Renormalization Group
(RG)~\cite{Wil}, \aka
the Exact Renormalization Group, has the potential to be an extremely
powerful framework for both exact and approximate calculations in
non-perturbative quantum field theory~\cite{devel}. 

However, a crucial challenge to confront has existed  since then: the
application to continuum gauge field theories. Since the introduction of
an effective cutoff breaks the gauge invariance, the Wilsonian approach
cannot be na\"\i vely followed. 

Following the earlier works of one of
us~\cite{Mor}, we have proposed a manifestly gauge invariant exact RG
equation, by which we can compute the effective action and extract the
physics, without ever requiring any gauge fixing or ghosts. The Polchinski
equation~\cite{Po} is modified by \eg introducing a Wilson line between 
functional derivatives, and thus it becomes gauge invariant.  

Needless to say, the theory is also to be regularised in a gauge invariant
way, and this is why a Poincar\'e and gauge invariant regularisation 
scheme has been worked out that is based on a real ultraviolet
scale.\footnote{as opposed to \eg analytic continuation of perturbative
amplitudes in dimensional regularisation.} It works for pure $SU(N)$
Yang-Mills theory  
in dimension four or less and, in the large $N$ limit, in any
dimension~\cite{us,morpro}.  

Such a regularisation scheme can be very nicely incorporated into our gauge
invariant formulation of the RG and, together, they constitute a very
powerful tool for investigating non-perturbative aspects of quantum
theories. Indeed, the use of manifest gauge invariance is the key ingredient
throughout the calculation.     

However, in view of the novelty of the present construction, it is
desirable to test the formalism first. We computed the one-loop beta
function for $SU(N)$ 
Yang-Mills theory for a general cutoff profile\footnote{provided some
general requirements on normalisation and ultraviolet decay rate are
satisfied} and we obtained the 
usual perturbative result, which is an encouraging confirmation that the
expected universality of the continuum limit has been incorporated.   
The calculation is completely independent of the details put in by hand,
\eg the choice of covariantisation and seed
action (which will be defined later in Section~\ref{sec:uno}), and
therefore guides us to a procedure for streamlined computations.

This note is organised as follows. In Section~\ref{sec:uno} we state the
flow equation in superfield notation, perform the usual loop
expansion and sketch our strategy for computing
$\beta_1$. Section~\ref{sec:gi} is devoted to listing the (un-)broken gauge
invariance identities, while 
Section~\ref{sec:due} contains a more detailed description
of the simplest part of the calculation, the scalar sector. Finally, in
section~\ref{sec:concl} we summarise and draw our conclusions,  
For further details on the regularisation scheme and notation, we refer the
reader to~\cite{morpro} and references therein.   
\section{$\boldsymbol{SU(N|N)}$ flow equation and its loop expansion}
\label{sec:uno}
Our manifestly gauge invariant exact RG equation in the unbroken phase,
\ie before shifting $\C \rightarrow \C + \sigma_3$, may
be written as\\ 
\begin{minipage}[t]{0.43 \linewidth}
\be
\ldl S = -  a_0[ S,\Sigma_g ] + a_1[\Sigma_g],
\ee
\ \\
\noindent
with $a_0[ S,\Sigma_g ]$ and $a_1[\Sigma_g]$ being the classical term and
quantum correction contributions respectively and $\Sigma_g = g^2 S - \hS$.
\end{minipage}
\vspace{-3.45cm}
\begin{figure}[h]
\hspace{.46 \linewidth}\ \begin{minipage}[t]{0.53 \linewidth}
\psfrag{=}{\hspace{-.7em}$=$}
\psfrag{-}{$-$}
\psfrag{ldl}{\hspace{-.7em}$\ldl$}
\psfrag{S}{$S$}
\psfrag{si}{$\Sigma$}
\psfrag{1/2}{  }
\psfrag{-1/l2}{\hspace{-.8em} ${\ds -\frac{1}{\Lambda^2}}$}
\psfrag{f}{ \hspace{-.4em}\tiny $f$} 
\psfrag{sumi}{\hspace{-.1em}${\displaystyle \sum_f}$}
\begin{flushright}
\epsfig{file=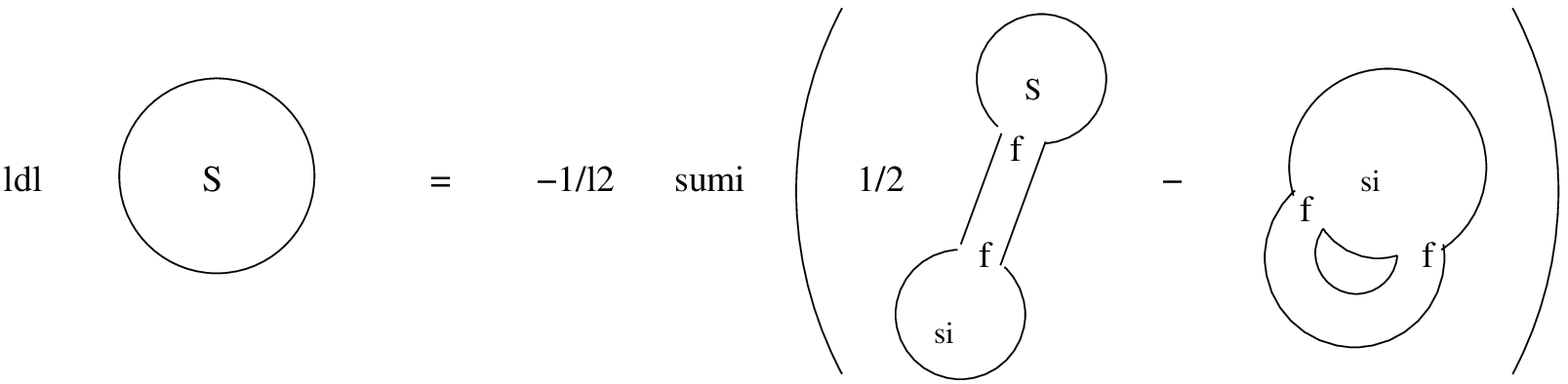, scale=.41}
\end{flushright}
\caption{Graphical representation of the flow equation.}\label{fig:floeq}
\end{minipage}
\end{figure}

$\sh$, hereafter referred to as the seed action, is part of the immense
freedom on the form of the RG equation~\cite{Lat, scalar, morpro}.
The explicit expressions for $a_{0,1}$ are given by     
\ba \label{a0a1}
a_0[S,\Sigma_g] &=& \frac{1}{2\Lam^2} \!\left( \!\frac{\delta
S}{\delta \A_{\mu}}\{c'\} \frac{\delta \Sigma_g}{\delta \A_{\mu}} 
- \frac{1}{4}[\C,\frac{\delta S}{\delta \A_{\mu}}]\{M\}[\C,\frac{\delta
\Sigma_g}{\delta \A_{\mu}}] \right) \nonumber\\
&&+ \frac{1}{2\Lam^4} \!\left( \!\frac{\delta S}{\delta
\C}\{H\}\frac{\delta \Sigma_g}{\delta \C} -\frac{1}{4}[\C,\frac{\delta
S}{\delta \C}]\{L\}[\C,\frac{\delta \Sigma_g}{\delta \C}] \right),
\nonumber\\
a_1[\Sigma_g] &=& \frac{1}{2\Lam^2} \!\left( \!\frac{\delta }{\delta
\A_{\mu}}\{c'\} \frac{\delta \Sigma_g}{\delta \A_{\mu}} -
\frac{1}{4}[\C,\frac{\delta }{\delta \A_{\mu}}]\{M\}[\C,\frac{\delta
\Sigma_g}{\delta \A_{\mu}}]  \right) \nonumber\\
&&+ \frac{1}{2\Lam^4} \!\left(\! \frac{\delta }{\delta \C}\{H\}\frac{\delta
\Sigma_g}{\delta \C} -\frac{1}{4}[\C,\frac{\delta }{\delta
\C}]\{L\}[\C, \frac{\delta \Sigma_g}{\delta \C}] \right),\nonumber\\
\ea 
where $\{W\}$ stands for any covariantisation of the kernel $W$. For any
two supermatrix representations, ${\bf u}, {\bf v}$, we can expand ${\bf
u}\{W\}{\bf v}$ in powers of the gauge field; the vertices of such an
expansion will be hereafter referred to as kernel's vertices (for
an example of how to covariantise $W$ and its vertices refer to \cite{Mor,
morpro}). 
  
The relations (\ref{a0a1}) differ from the usual Polchinski-type equation
by the presence of terms involving commutators.\footnote{The commutators
guarantee 
the equation is still gauge invariant.} These latter, once the theory is
spontaneously broken, will affect the equations for the two-point bosonic
and fermionic vertices differently, and by choosing the related kernels
properly, all the two-point vertices of
the effective action can be set equal to the hatted ones (cf. \cite{morpro,
us2}). This last requirement will greatly simplify the equations for the higher
order interactions, and thus will be enforced for convenience. 

The resulting kernels in \eq{a0a1} are found to be
\be
M(x) = - \left( \frac{2 c^2}{x \tilde{c} + 2 c}
\right)'\hspace{-.5em}, \quad x H(x) =  \left( \frac{2 x^2 \tilde{c}}{x + 2
\lambda \tilde{c}} \right)'\hspace{-.5em}, \quad
x L(x) = \left( \frac{x^2 \tilde{c} (\lambda \tilde{c}^2 - c)}{(x + 2 \lambda
\tilde{c}) (x \tilde{c} + 2 c) } \right)'\hspace{-.5em}, 
\ee
where prime denotes differentiation with respect to $x$ and $c, \tilde{c}$
are meant to be functions of $x$.

\begin{figure}[h]
\begin{minipage}[t]{0.38 \linewidth}
\psfrag{sim}{\hspace{-.2em}$=$}
\psfrag{=}{\tiny $=$}
\psfrag{Wp}{$c'_p$}
\psfrag{Kp}{\hspace{-.1em}$K_p$}
\psfrag{Hp}{$H_p$}
\psfrag{Gp}{$G_p$}
\psfrag{Am}{$A_{\mu}$}
\psfrag{Bm}{$B_{\mu}$}
\psfrag{C}{$C$}
\psfrag{D}{$D$}
\begin{flushleft}
\includegraphics[scale=.33]{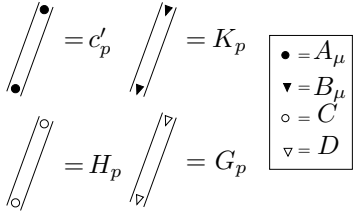}
\end{flushleft}
\caption{Graphical representation of 0-point kernels. The $f$-kernel in
Fig.~\ref{fig:floeq} stands for any of these. $K \doteq c'+M$.}\label{fig:0wines}
\end{minipage}
\end{figure}

\begin{figure}[h]
\hspace{.4 \linewidth} \begin{minipage}[t]{0.6 \linewidth}
\psfrag{pm}{\tiny $p_{\mu}$}
\psfrag{qa}{\tiny $q_{\alpha}$}
\psfrag{rb}{\tiny $r_{\beta}$}
\psfrag{Hm}{$\frac{1}{\Lam^2}H_{\mu}('')$}
\psfrag{=}{\hspace{-.3em}$=$}
\psfrag{cpm}{$c^{\prime}_{\mu}(p;q,r)$}
\psfrag{Km}{$K_{\mu}('')$}
\psfrag{Gm}{$\frac{1}{\Lam^2}G_{\mu}('')$}
\psfrag{m1}{$\frac{M_q+M_r}{2}$}
\psfrag{m2}{$\frac{M_q}{2}$}
\psfrag{m3}{$\frac{M_r}{2}$}
\begin{center}
\includegraphics[scale=.3]{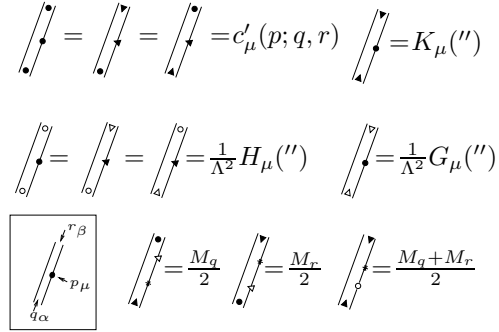}
\end{center}
\caption{Graphical representation of 1-point kernels. The boxed diagram
indicates the position of incoming momenta. The $\sigma_3$s coming from
the symmetry breaking are represented by stars, while $(``)$ stands for
$(p;q,r)$.
}\label{fig:1wines}
\end{minipage}
\end{figure}

Breaking supersymmetry spontaneously in the fermionic directions causes
the fermionic fields to acquire a mass of order the effective cutoff,
whilst leaving the physical gauge field and its copy massless. In the unitary
gauge, the Goldstone modes $D$ vanish (eaten by the $B$'s), leaving the
massive ``Higgs'' $C$ and the massive vector fermions, $B_\mu$.  
The symmetries thus suggest that superfields be split into
their diagonal and off-diagonal components, \ie $\A_\mu = A_\mu + B_\mu$ and
$\C = C+D$ \cite{morpro,
us2}. This makes the equations much easier to derive and, also,
resembles what is usually done in the context of the standard model, namely
to deal with massive and massless combination of gauge fields rather than
with the fields themselves. The kernels and their expansion in powers of
gauge fields may be split accordingly. (See Figs.~\ref{fig:0wines} and
\ref{fig:1wines} for some of those actually used in the calculation.)


\vspace{-5.6cm}
  
Expanding the action and the beta function $\beta(g) = \ldl g$ in powers
of the coupling constant:
\be
S \ug\frac{1}{g^2}S_0+S_1+g^2 S_2+\cdots \qquad \quad \beta \ug
\Lambda\partial_{\Lambda}g=\beta_1g^3+\beta_2g^5+\cdots 
\ee
yields the loopwise expansion of the flow equation
\ba
\Lambda\partial_{\Lambda}S_0 &\ug& -a_0[S_0,S_0-2\hat{S}],\label{treelevelfloeq}\\
\Lambda\partial_{\Lambda}S_1 &\ug& 2\beta_1 S_0-2a_0[S_0-\hat{S},S_1]+
a_1[S_0-2\hat{S}]\label{oneloopfloeq},
\ea
{\it etc.}, where $S_0$ ($S_1$) is the classical (one-loop) effective
action.
The one-loop coefficient, $\beta_1$, can be extracted directly
from \eq{oneloopfloeq} once the renormalization conditions are imposed. Since
gauge invariance already forces the anomalous dimension of the gauge field
to vanish~\cite{Mor,morpro,us2}, we only need to define the
renormalized coupling $g(\Lam)$. This is done via the field strength for
the physical gauge field, $A^1_\mu$ \cite{Mor,morpro,us2},  
\be \label{rencon}
S = \frac{1}{2 g^2(\Lam)} \tr \int \dd^D x \, (
F_{\mu\nu}^1 )^2  +{\cal O}(\nabla^3).
\ee

From the above equation
\be \label{renc}
S_{\mu \nu}^{AA}(p) +  S_{\mu \nu}^{AA\sigma}(p) = \frac{2}{g^2}
\Box_{\mu \nu} (p) + {\cal O}(p^3) = \frac{1}{g^2} S_0 {}_{\mu
\nu}^{AA}(p)+ {\cal O}(p^3), 
\ee
\noindent 
with  $\Box_{\mu \nu} (p)$ being the transverse combination $(p^2 \delta_{\mu \nu} -p_\mu p_\nu)$.
Eq.~(\ref{renc}) implies the ${\cal O}(p^2)$  component of all the higher loop
contributions $
S_n{}_{\mu \nu}^{AA}(p) +  S_n{}_{\mu \nu}^{AA\sigma}(p)$ must vanish. Thus
the equation for $\beta_1$ becomes algebraic ($\Sigma_0 = S_0 -2 \hat{S}$):
\be \label{beta1}
-2 \beta_1 S_0{}_{\mu \nu}^{AA}(p) + {\cal O}(p^3) = a_1[\Sigma_0]_{\mu
 \nu}^{AA}(p). 
\ee

\psfrag{=}{\hspace{1.2em} $=$}
\psfrag{mu}{\small $\mu$}
\psfrag{nu}{\small $\nu$}
\psfrag{Si}{\hspace{-.3em}$\Sigma_0$}
\psfrag{F}{\tiny $f$}
\psfrag{+}{\hspace{.1em}$+$}
\psfrag{-}{$-$}
\psfrag{S0}{$S_0$}
\psfrag{ldl}{$\Lambda\partial_{\Lambda}$}
\psfrag{Sum}{  \hspace{.7em}${\ds \sum_f}$}
\psfrag{1}{  $1$}
\psfrag{b}{\hspace{-3.6em}$-4\beta_1 \Box_{\mu \nu} (p)$}
\psfrag{O(p3)}{ \hspace{.3em}$O(p^3)$}
\psfrag{2/l2}{}
\begin{figure}[h]
\begin{center}
\includegraphics[scale=.3]{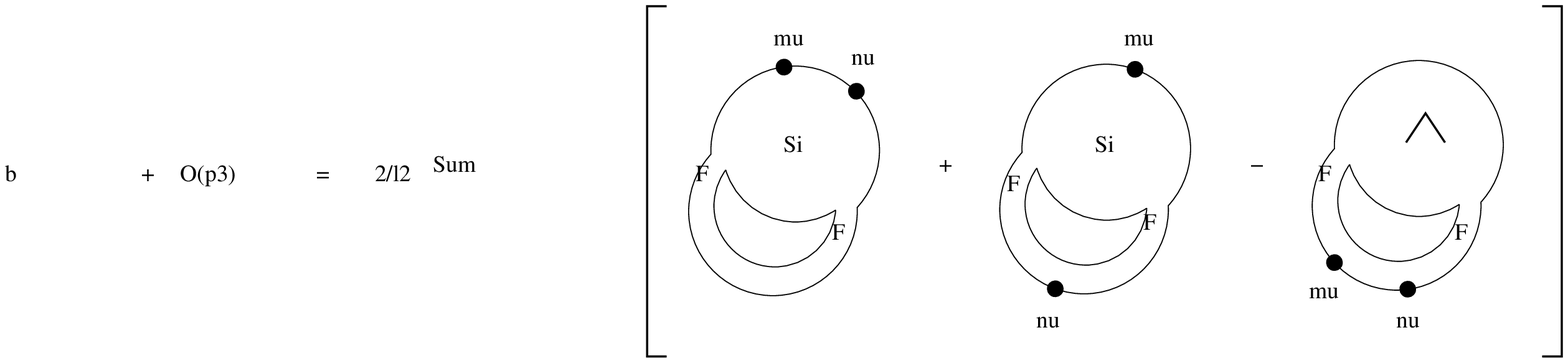} 
\caption{Graphical representation of \eq{beta1}.}\label{fig:beta1}
\end{center}
\end{figure}

In order to calculate the r.h.s. of \eq{beta1}, we will adopt the following
strategy:\\ 
i.\phantom{ii} introduce the ``integrated kernels'' in the $S_0$ part
of the first diagram and integrate by parts so as to end up with
$\Lam$-derivatives of vertices of the effective action;\\
ii.\phantom{i} use the equations of motion for the effective
couplings;\\
iii. use the relation between the integrated kernels and their
corresponding two-point functions to simplify the diagrams obtained so far;\\
iv.\phantom{i} repeat the above procedure when any three-point
effective coupling is generated.

This simple procedure, which will be described in more detail in the next
section, ensures that any dependence upon $n$-point vertices of the seed
action, $n \geq 3$, will cancel out. This implies that the calculation is
actually independent of the choice of $\hat{S}$, provided it is a
covariantisation of its two-point vertices\footnote{set equal to the
effective ones for convenience.} and these latter vertices
are infinitely differentiable and lead to convergent momentum
integrals~\cite{scalar,antonio,morpro}.
Moreover, pursuing that strategy will also guarantee that just the kernels'
vertices with special momenta remain that by gauge
invariance can be expressed as derivatives of their generators (for an
example see Section~\ref{sec:gi}), which means
independence of the choice of covariantisation.  
\section{(Un-)Broken gauge invariance}
\label{sec:gi}
The invariance under the (broken) $SU(N|N)$ gauge symmetry results in the
following set of trivial Ward identities 
\be\label{gi}
\bea
q^{\nu}U^{\cdots X A Y\cdots}_{\cdots \,a \,\, \nu \,\, b\cdots}(\cdots
p,q,r,\cdots)=U^{\cdots X Y\cdots}_{\cdots \,a \,\,\,  b\cdots}(\cdots
p,q+r,\cdots)
-U^{\cdots X Y\cdots}_{\cdots \,a \,\,\, b\cdots}(\cdots
p+q,r,\cdots),\\[4pt]
q^{\nu}U^{\cdots X B Y\cdots}_{\cdots \,a\,\, \nu\,\, b\cdots}(\cdots
p,q,r,\cdots)=\pm U^{\cdots X \hat{Y}\cdots}_{\cdots \,a\,\,\,  b\cdots}(\cdots
p,q+r,\cdots) \mp U^{\cdots \hat{X} Y\cdots}_{\cdots \,a\,\,\,  b\cdots}(\cdots
p+q,r,\cdots) \\[4pt]
\hspace{12.7em}+2 U^{\cdots X D\sigma Y\cdots}_{\cdots \,a \phantom{D\sigma}\,\,b\cdots}(\cdots
p,q,r,\cdots),\\
\eea
\ee 
where $U$ is any vertex, $a$ and $b$ are Lorentz indices or null as
appropriate and $\hat{X},\hat{Y}$ are opposite statistics partners of
$X,Y$. The sign of the terms containing $\hat{X},\hat{Y}$ depends on whether
$B$ goes past a $\sigma_3$ on its way back and forth. 

By specialising (\ref{gi}) to a proper set of momenta, one of which has to
be infinitesimal, it 
is possible to express $n$-point vertices with one null momentum as
derivatives of $(n-1)$-point's, independently of the choice of
covariantisation. As an example, let us consider the three-point pure-$A$
effective vertex at vanishing first momentum, $S^{AAA}_{\,\mu \,\nu
\,\rho}(0, k, -k)$. By using (\ref{gi}), 
\be
\epsilon^\mu S^{AAA}_{\,\mu \,\nu \,\rho}(\epsilon, k, -k-\epsilon) =
S^{AA}_{\,\nu \,
\rho}(k+\epsilon) - S^{AA}_{\,\nu \,\rho}(k)
= \epsilon^\mu \, \demu^k \, S^{AA}_{\,\nu\, \rho}(k) + {\cal O}(\epsilon^2).
\ee
At order $\epsilon$, $S^{AAA}_{\,\mu\, \nu\, \rho}(0, k, -k) = \demu^k \,
S^{AA}_{\,\nu\, \rho}(k)$.
Also $S^{AAAA}_{\,\mu\, \nu \,\rho \,\sigma}(0, 0, k, -k) = {1\over 2} \demu^k \,
\denu^k \,S^{AA}_{\,\rho \,\sigma}(k)$. 

\section{A sample of the calculation: the $\boldsymbol{C}$ sector}
\label{sec:due}
In this section the simplest part of the computation will be described,
that is the scalar sector. All the steps of the strategy previously outlined  
will be illustrated by means of diagrams, as the cancellations taking place
are  
evident already at that level. Of course, performing the full and complete
calculation yields the same result.

We start by defining the integrated kernel. As $\ldl f(\frac{p^2}{\Lam^2})
= -2 \frac{p^2}{\Lam^2} f'(\frac{p^2}{\Lam^2})$,\\
\begin{minipage}[t]{0.53 \linewidth}
\be \label{intker}
\frac{1}{\Lam^4} H = -\frac{1}{2 p^4}
\ldl \left(\frac{2 x^2 \tilde{c}}{x+2 \lambda
\tilde{c}} \right) = - \ldl \Delta^{CC} 
\ee
\end{minipage}
\vspace{-1.4cm}
\begin{figure}[h]
\hspace{.56 \linewidth}\ \begin{minipage}[t]{0.42 \linewidth}
\begin{flushright}
\psfrag{=}{$=$}
\psfrag{-}{$-$}
\psfrag{ldl}{$\ldl$}
\includegraphics[scale=.5]{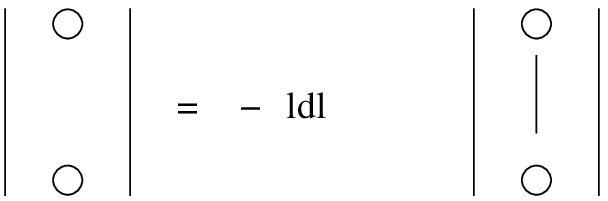}
\end{flushright}
\end{minipage}
\end{figure}

\noindent
\hspace{.58 \linewidth}\ \begin{minipage}[t]{0.4 \linewidth} 
The integrated kernel is introduced via \eq{intker} into the $S_0$ part of
the first diagram in Fig.~\ref{fig:beta1}. One then integrates
by parts, so as to end up with a total $\Lam$-derivative plus the
tree-level $\ldl S^{AACC}_{\mu \, \nu}$ vertex joined by a
$\Delta^{CC}$. The latter will be dealt with, using the equations of
motion.    
\end{minipage}
\vspace{-3.4cm}
\begin{figure}[h!]
\begin{minipage}[t]{0.6 \linewidth}
\vspace{-\abovedisplayskip}
\begin{flushleft}
\psfrag{=}{$=$}
\psfrag{S0}{\small 0}
\psfrag{-}{$-$}
\psfrag{ldl}{\hspace{-.35em}$\ldl$}
\includegraphics[scale=.3]{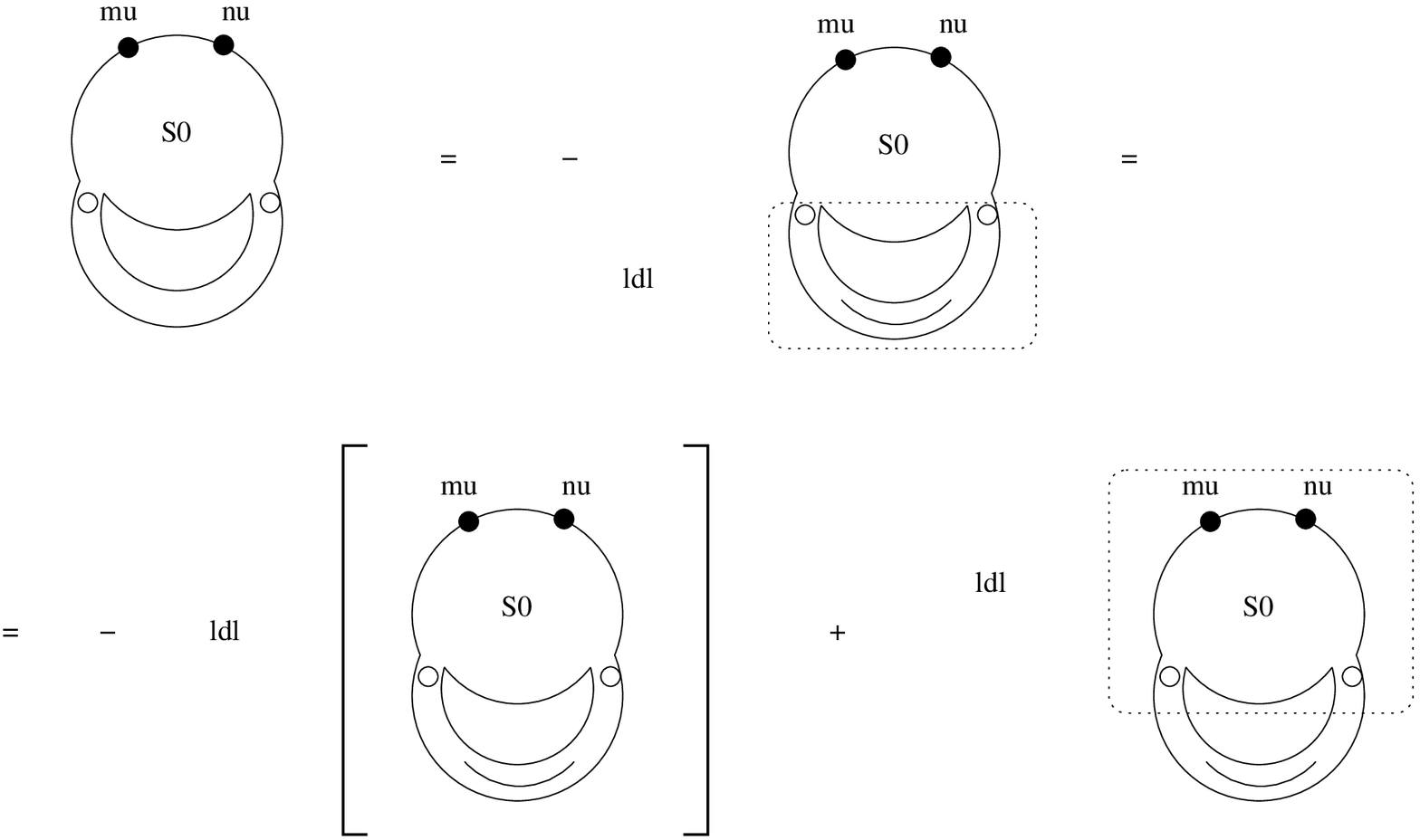}
\end{flushleft}
\end{minipage}
\end{figure}

\begin{figure}[h!]
\begin{center}
\psfrag{=}{$=$}
\psfrag{-}{$-$}
\psfrag{+}{$+$}
\psfrag{mu}{\small $\mu$}
\psfrag{nu}{\small $\nu$}
\psfrag{S0}{\hspace{.2em}\tiny $0$}
\psfrag{cdots}{$\cdots$}
\psfrag{ldl}{$\ldl$}
\includegraphics[scale=.4]{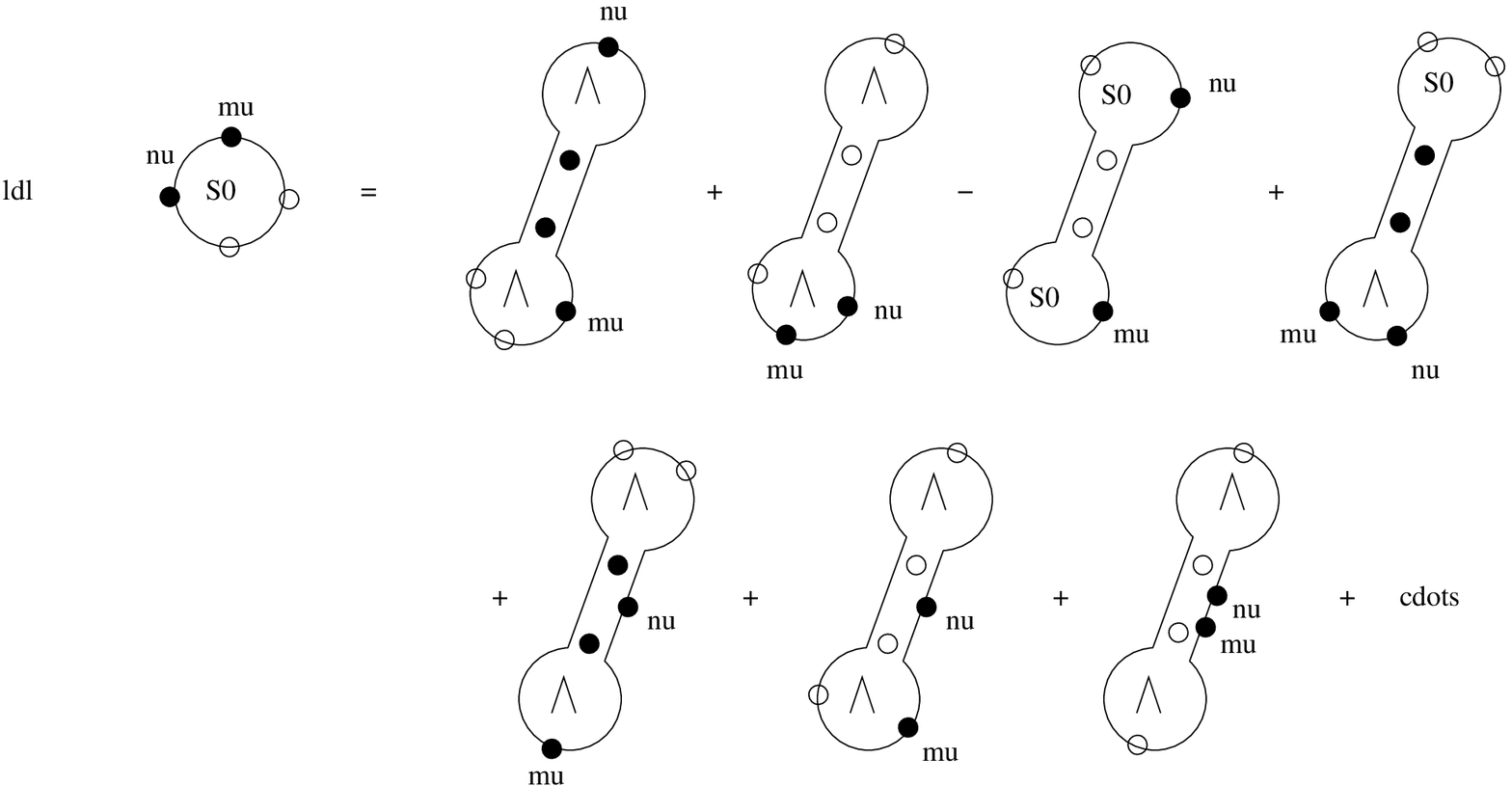}
\caption{Eq.~(\ref{treelevelfloeq}) as specialised to 
$S^{AACC}_{\mu \, \nu}$. The ellipsis stands for similar diagrams which
have not been drawn.}\label{fig:saacc}    
\end{center}
\end{figure}

The next step consists in using \eq{treelevelfloeq} as specialised to 
$S^{AACC}_{\mu \, \nu}$. Some of the diagrams are shown in
Fig.~\ref{fig:saacc}.  

\begin{figure}[h!]
 \begin{minipage}[t]{0.43 \linewidth}
\begin{center}
\psfrag{mu}{\small $\mu$}
\psfrag{nu}{\small $\nu$}
\includegraphics[scale=.35]{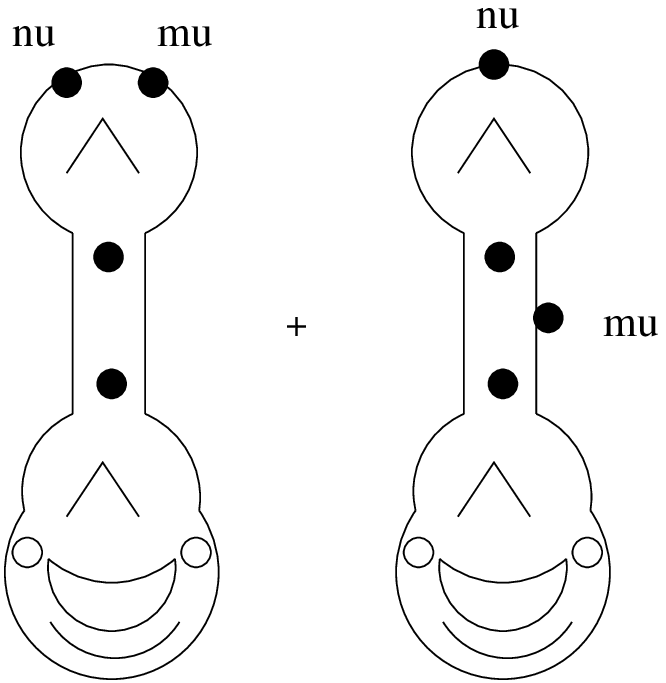}
\caption{Diagrams not contributing to $\beta_1$.}\label{fig:three}
\end{center}
\end{minipage}
\end{figure}

\begin{figure}[h!]
 \begin{minipage}[t]{0.43 \linewidth}
\begin{center}
\psfrag{mu}{\small $\mu$}
\psfrag{nu}{\small $\nu$}
\includegraphics[scale=.35]{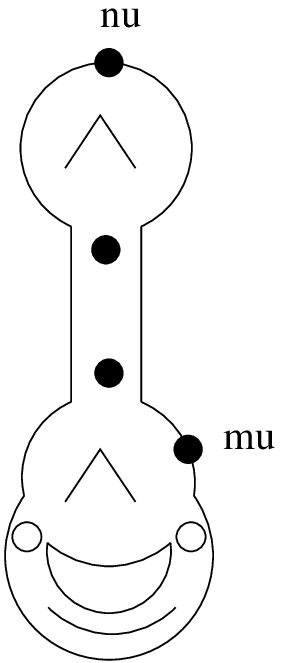}
\caption{A potentially universal contribution.}\label{fig:one}
\end{center}
\end{minipage}
\end{figure}

Already at this level, we note that some of the diagrams either do not
contribute 
at all (cf. Fig.~\ref{fig:three}) or they give a potentially universal
contribution, \ie something depending only on two-point vertices and
integrated kernels (cf. Fig.~\ref{fig:one}).

\noindent
\hspace{.43 \linewidth}\ \begin{minipage}[t]{0.57\linewidth} 
$$
\bea
\Big(\sh^{AAA}_{\mu \nu \alpha}(p,-p,0) \, c'(0) \,
\sh^{ACC}_{\alpha}(0,k,-k) +\\[0.3cm]
\left. \sh^{AA}_{\nu \alpha}(p) \, c'_\mu(p;0,-p) \,
\sh^{ACC}_{\alpha}(0,k,-k) \Big) \Delta^{CC}(k) \right|_{p^2}=0
\eea
$$   
\end{minipage}
\vspace{-2.5cm}

\noindent
\hspace{.43 \linewidth}\ \begin{minipage}[t]{0.57\linewidth} 
$$
\bea
\left.\sh^{AA}_{\nu \alpha}(p) \, c'(\textstyle{p^2\over \Lam^2}) \,
\sh^{AACC}_{\mu \alpha}(p,-p,k,-k) \, \Delta^{CC}(k) \right|_{p^2}=\\[0.3cm]
\sh^{AA}_{\nu \alpha}(p) \, c'(0) \, \sh^{AACC}_{\mu
\alpha}(0,0,k,-k) \, \Delta^{CC}(k)=\\[0.3cm]
{1\over 2} \sh^{AA}_{\nu \alpha}(p) \, c'(0) \, \demu^k \partial_\alpha^k
\sh^{CC}(k) \, \Delta^{CC}(k). 
\eea
$$   
\end{minipage}
\vspace{-2.7cm}

\begin{figure}[h!]
\begin{center}
\psfrag{2}{}
\psfrag{mu}{\small $\mu$}
\psfrag{nu}{\small $\nu$}
\psfrag{ug}{$\parallel$}
\includegraphics[scale=.45]{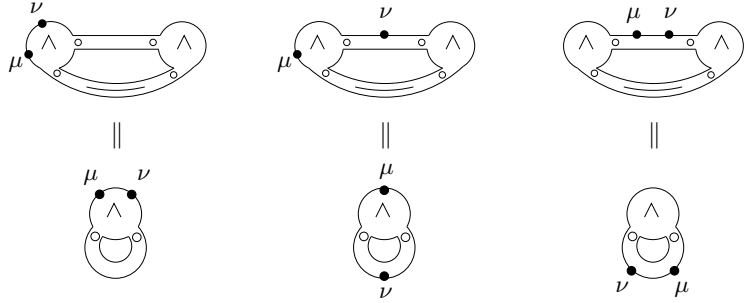}
\caption{Simplifications in the four-point effective vertex
contribution.}\label{fig:simpl} 
\end{center}
\end{figure}

Many of the remaining terms in the tree-level equation for $S^{AACC}_{\mu
\, \nu}$ may be further 
simplified by making use of the relation between the integrated kernel and
the corresponding two-point function. Such a relation may be easily
obtained from the 
tree-level equation for the effective two-point coupling, in the present
example $S^{CC}$. By rewriting it in terms of the inverse coupling,
$(S^{CC})^{-1}$, we get $(S^{CC})^{-1} = \Delta^{CC}$, \ie $S^{CC} \,
\Delta^{CC}=1$. This leads to the simplifications shown in
Fig.~\ref{fig:simpl}.  

\begin{figure}[!h]
\psfrag{0}{\tiny 0}
\psfrag{mu}{\small $\mu$}
\psfrag{nu}{\small $\nu$}
\psfrag{=}{\small $=$}
\psfrag{1/2}{$\frac{1}{2}$}
\psfrag{cdots}{$cdots$}
\psfrag{ldl}{\small \hspace{-.3em}$\ldl$}
\begin{center}
\includegraphics[scale=.35]{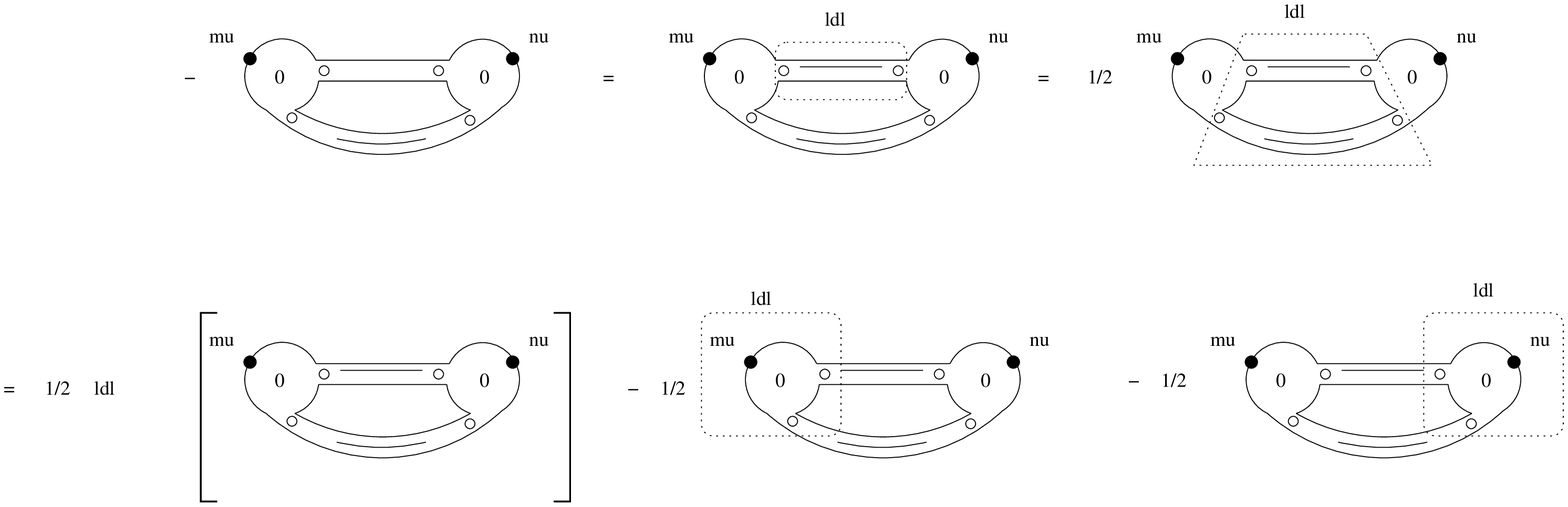}
\caption{How to handle two joined three-point effective
vertices.}\label{fig:trick2}  
\end{center}
\end{figure}

The last step concerns how to handle the terms that contain two three-point
effective couplings. The procedure is pretty much the same, except that one
has to recognise the derivative of the ``square of the kernel'' (see
Fig.~\ref{fig:trick2}). At the
algebra level, it amounts to writing the second diagram in
Fig.~\ref{fig:trick2} 
as the sum of two equal contributions and, then, to shifting the loop
momentum so as to complete the $\Lam$-derivative. 
\noindent
\noindent

The procedure outlined in the above can be used in the whole
calculation: all the hatted vertices cancel out and one is left with
potentially universal terms only. The relation
between integrated kernels and their corresponding two-point functions,
however, is more complicated in the general case. As a matter of fact, it
takes the form 
$S_{IK}(p) \Delta_{KJ}(p) = \delta_{IJ} + R_{IJ}(p)$,
where the ``remainder'' $R_{IJ}$, absent in the scalar sector, is a
(un-)broken gauge transformation. In the $A$ sector, for example,
$R_{\mu \nu}(p) = - \frac{p_\mu \, p_\nu}{p^2}$. 

Once the potentially universal terms have been collected, the momentum
integrals should be carried out. We used dimensional
regularisation as a preregulator to avoid all the subtleties related to
cancelling divergences against each other. (Had we done the calculation 
in a way that preserves $SU(N|N)$, preregularisation would not have been
needed.)   
\section{Conclusions}
\label{sec:concl}
 
A manifestly gauge invariant RG has been proposed. Together with
the necessary gauge invariant regularisation~\cite{us}, it constitutes a
very powerful 
method of computation in gauge theory, as it allows one to
calculate the Wilsonian effective action and extract the physics in a way
that respects the gauge symmetry, with no need for gauge fixing and 
accompanying ghosts~\cite{Mor, us2}. Hence the Gribov problem is completely
avoided~\cite{gribov}.  

As a basic test of the formalism, the one-loop $SU(N)$ beta function has
been computed and the expected universal result has been obtained. 
The strategy which has proven to be very efficient consists in eliminating
the elements put in by hand by using the equations of motion for the
effective action vertices, where physics is actually encoded. (See also
\cite{scalar} for the analysis of the scalar case).  
A diagrammatic technique to represent the various vertices has been
sketched, and already at the level of diagrams the big potential of the
method comes out.

The calculation is totally independent of the details put in by hand, 
such as the choice of covariantisation and the cutoff profile, and gauge
invariance is no doubt the main ingredient all the way to the final result. 

We hope the procedure is quite general and may be used to investigate
non-perturbative aspects of gauge theories. 

Including matter in the fundamental representation as well as space-time
supersymmetry seems not too difficult, thus opening the door to many
further avenues of exploration.

\begin{ack}
The authors wish to thank the organisers of the fifth international
Conference ``RG 2002'' for providing such a stimulating environment. 
T.R.M. and S.A. acknowledge financial support from PPARC Rolling Grant
PPA/G/O/2000/00464.   
\end{ack}

\end{document}